 \documentclass[letter,twocolumn]{jpsj2}

\def\runauthor{Tota Nakamura}

\title{%
Dynamic crossover in the spin-glass phase
}

\author{%
Tota Nakamura
}

\inst{%
Laboratory of Natural Science, Shibaura Institute of Technology\\
307 Fukasaku, Minuma, Saitama 337-8570
}

\recdate{\today}

\abst{%
Dynamic scaling analyses are performed in the spin-glass
phase of the $\pm J$ Ising, the {\it XY}, and the Heisenberg models 
in three dimensions.
We found a crossover from the critical dynamics to the ground-state dynamics
in the Ising model and the Heisenberg model.
The ground-state dynamics of the Ising model is characterized by an activation
law with a finite energy gap: the typical time diverges exponentially.
On the other hand,
the typical time in the Heisenberg model diverges algebraically with
the inverse temperature.
Algebraic relaxation with a finite dynamic exponent is observed after the
typical time in both models.
The ground-state dynamic exponent is estimated to be
$z_0\simeq 13$, which is common to both models.
There is no crossover in the {\it XY} model.
The critical dynamics is considered to continue to the ground-state.
}

\kword{%
spin glass phase, dynamic scaling, dynamic exponent, droplet picture, 
RSB picture
}

\begin{document}
\maketitle

Randomness and frustration produce nontrivial phenomena in
spin glasses\cite{SGReview3}.
Theoretical investigations in spin glasses are mostly concentrated into two
topics.
One is an issue regarding a picture of the spin-glass (SG) phase.
There have been many arguments
whether it is explained by the replica-symmetry-breaking(RSB) theory 
\cite{RSB} or it is explained by the droplet theory.\cite{droplet}
The second topic is an argument of existence or nonexistence of a
spin-glass transition in short-range Edwards-Anderson(EA) models.
There is a consensus that the SG transition occurs in the Ising 
model in three dimensions.\cite{Bhatt,maricampbell}
However, the issue is not settled on the vector spin models.
A chirality mechanism insists that a chiral-glass(CG) transition occurs without
the SG transition.\cite{KawamuraXY3,HukushimaH2}
However, there are several investigations suggesting a simultaneous 
transition.\cite{MatsubaraEndoh,nakamura,Lee,yamamoto1}

In our recent paper\cite{nakamuraSGzT}, 
the CG and the SG transition temperatures, $T_\mathrm{sg}$ and $T_\mathrm{cg}$,
are resolved to be different in the Heisenberg model.
We have performed a dynamic scaling analysis of the glass susceptibilities.
Particularly, we paid attention to a temperature dependence 
of the dynamic exponent, $z(T)$.
Since the exponent is a key variable in the dynamic scaling hypothesis,
a careful determination of $z(T)$ increases accuracy of the scaling results.

It is known that the dynamic exponent is proportional to the 
inverse temperature in the spin-glass phase.\cite{marinarizT,kisker1,komori1}
\begin{equation}
\frac{1}{z(T)} = \frac{T}{z_\mathrm{c} T_\mathrm{c}}.
\label{eq:zne}
\end  {equation}
The behavior is also observed experimentally.\cite{joh}
It has been considered an outcome of the slow dynamics in the spin-glass phase.
However, it is found recently \cite{katzgraber}
that the temperature dependence also appears
in the paramagnetic phase and there is no anomaly at the transition
temperature.
It is certified to be a general feature of the spin-glass models in 
three dimensions.\cite{nakamuraSGzT}

It is important to investigate whether or not the temperature dependence 
of the dynamic exponent continues to the zero temperature.
If it continues, a dynamics of the ground state is smoothly connected 
to the paramagnetic state.
We must consider a theory of the spin-glass dynamics including both
the ground state and the paramagnetic state.
If it does not continue and there is a crossover temperature, then
we may consider the higher and the lower temperature regions separately.
The ground-state (or close to the ground-state) dynamics is intrinsically
different from the critical dynamics near the spin-glass transition temperature.
Hukushima {\it et al.} \cite{huku00}
reported such a crossover in the four-dimensional $\pm J$ Ising model.

In this paper, we study the low-temperature phase of the
$\pm J$ Ising, the {\it XY}, and the Heisenberg models in three dimensions.
A dynamic scaling analysis is performed on the spin-glass correlation length.
The temperature dependences of the dynamic exponent are obtained.
We find a dynamic crossover in the Ising model and in the Heisenberg model.
The spin dynamics is essentially different above and below the crossover
temperature, $T_\mathrm{cr}$.

We consider an EA model on a simple cubic lattice.
The Hamiltonian is written as
\begin{equation} 
{\cal H}=\sum_{\langle i,j\rangle}J_{ij}
\mbox{\boldmath $S$}_i\cdot
\mbox{\boldmath $S$}_j.
\end  {equation} 
The interactions, $J_{ij}$, take the two values  $+J$ and $-J$
with the same probability.
The temperature, $T$, is scaled by $J$.
Linear lattice size is denoted by $L$. 
A total number of spins is $N = L \times L \times (L+1)$, and
skewed periodic boundary conditions are imposed.
The spins are updated by the single-spin-flip algorithm
using the Metropolis probability.

We start simulations with random spin configurations ($T=\infty$).
The temperature is quenched to a finite value at the first Monte Carlo step.
Physical quantities we observe are the spin-glass (chiral-glass) correlation
length, which is estimated from the correlation functions by a
single exponential fitting.
The relaxation functions are averaged over independent Monte Carlo simulations.

A typical lattice size in the Ising simulations is 71.
A replica number is 128, and typical numbers of random bond samples are
more than one hundred.
In the {\it XY} simulations,
the lattice size is 55, and the replica number is 64. 
The sample numbers are mostly around one hundred.
In the Heisenberg simulations,
the lattice size is 59, and the replica number is 32 or 64.
The sample numbers are between 10 and 100.

\begin{figure}[t]
\begin{center}
\includegraphics[width=7cm]{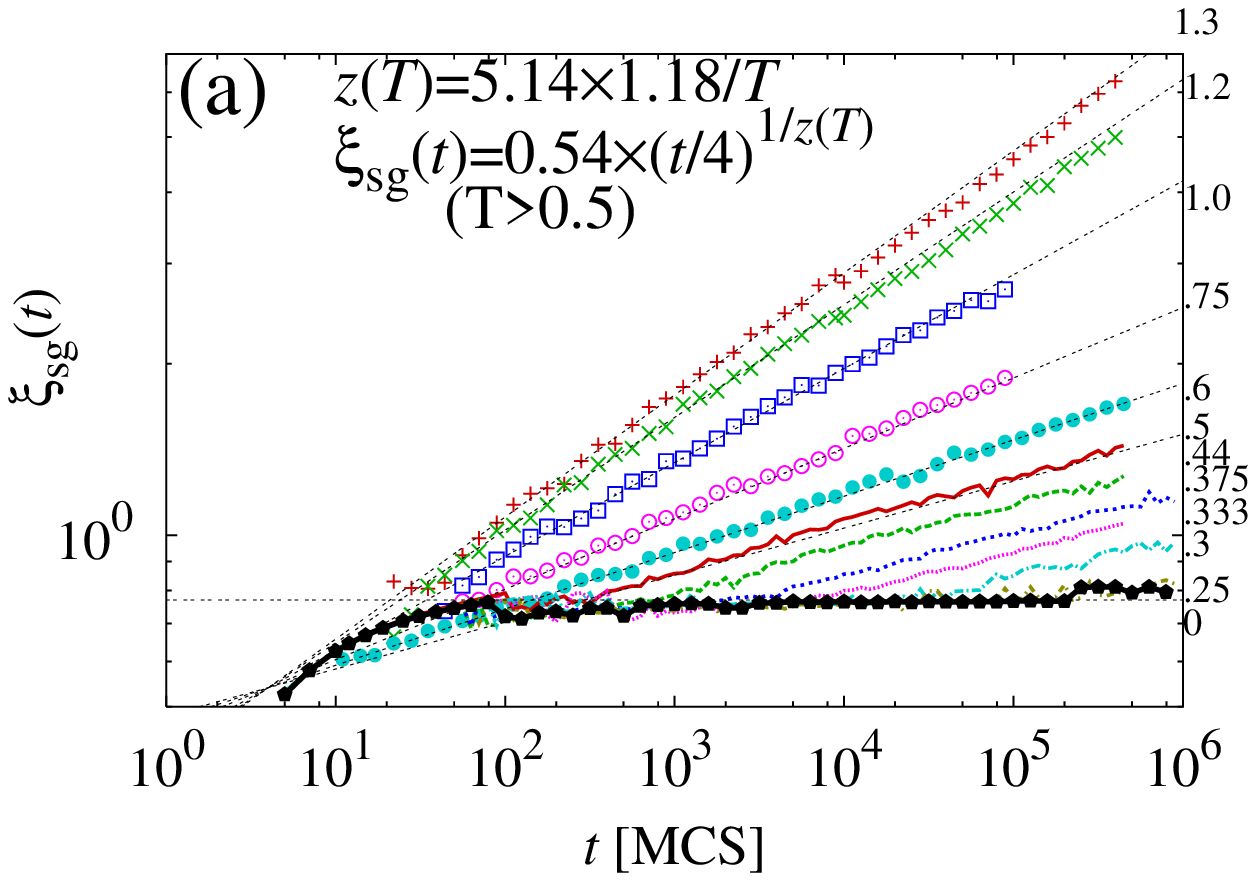}
\includegraphics[width=7cm]{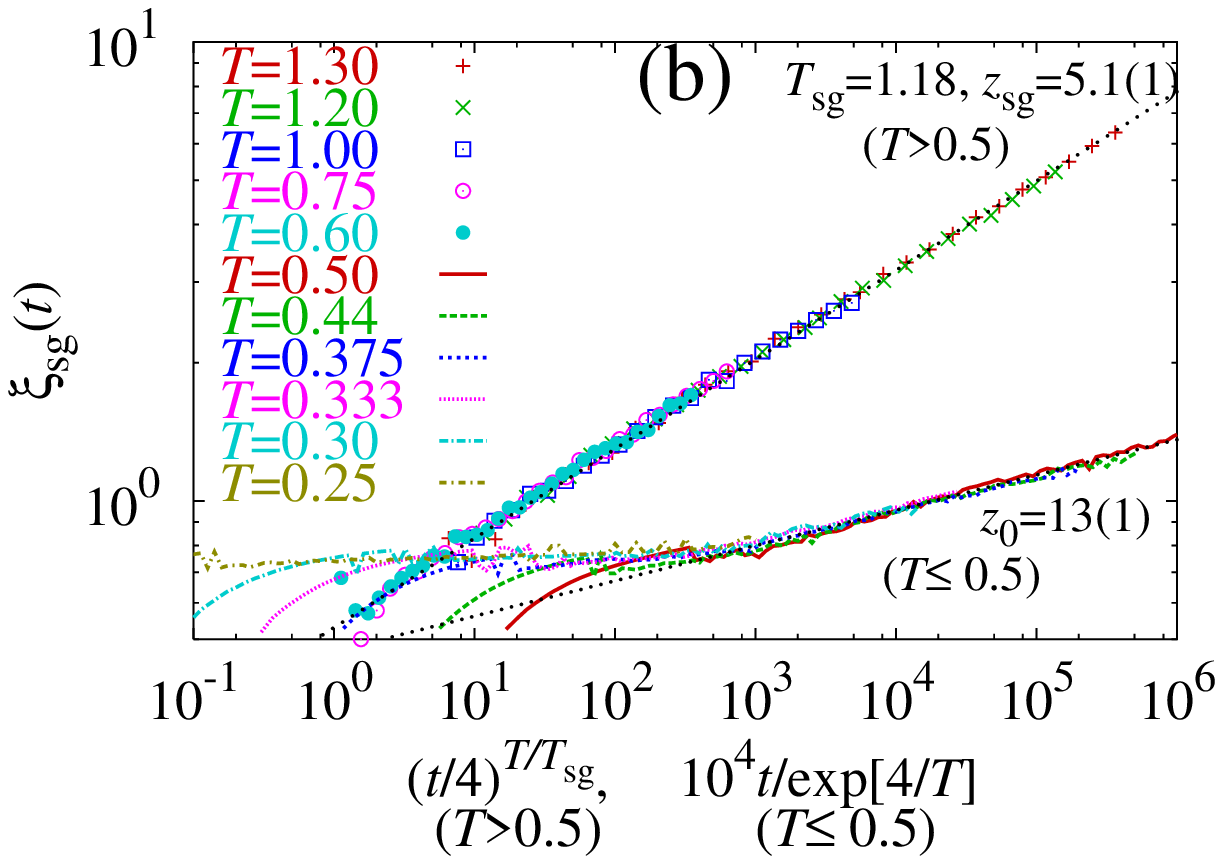}
\end{center}
\caption{(Color online)
(a) Nonequilibrium relaxation functions of the spin-glass correlation 
length in the Ising model.
The temperatures of data are denoted aside the fitting lines.
A relaxation function at $T=0$ is plotted by circle symbols.
(b)
The dynamic scaling plot.
When $T>T_\mathrm{cr}=0.5$, data are plotted
with symbols against $(t/4)^{T/T_\mathrm{sg}}$.
Below $T_\mathrm{cr}$, they are plotted with lines against
$10^4t/\exp[4/T]$.
}
\label{fig:isingxi}
\end{figure}

Figure \ref{fig:isingxi}(a) shows relaxation functions of the spin-glass 
correlation length in the Ising model.
They show algebraically diverging behaviors.
We estimate the nonequilibrium dynamic exponent, $z$, from the logarithmic
slope, since 
\begin{equation}
\xi(t) =\xi_0\times (t/\tau)^{1/z}
\label{eq:dynsca}
\end{equation}
from the dynamic scaling relation.
The dynamic exponent smoothly depends on the temperature 
when $T>T_\mathrm{cr}=0.5$.

The relaxation function changes the behavior at $T_\mathrm{cr}$.
It shows a plateau below this temperature.
As the temperature decreases, the plateau range becomes longer.
A spin state is considered to be trapped in a local energy minimum.
After some time steps, it eventually escapes from the valley and shows
an algebraic divergence, whose dynamic exponent is independent from the
temperature.
This is a distinct difference from the higher temperature region.
A temperature effect only appears in the dynamics of escaping from the
valley.
A dynamics after the escape is independent from the temperature.
It is natural to consider that this temperature-independent dynamics 
continues to the ground state.
Therefore, this is a crossover from the critical dynamics around the transition 
temperature to the ground state dynamics.

We perform a dynamic scaling analysis on the spin-glass correlation length.
A basic scaling form is the dynamic scaling relation, Eq.~(\ref{eq:dynsca}).
When $T>T_\mathrm{cr}$, a typical time, $\tau$, is a constant and the dynamic
exponent, $z$, is proportional to the inverse temperature.
The scaling analysis is performed by plotting $\xi(t)$ against 
$(t/\tau)^{T/T_\mathrm{sg}}$.\cite{bert1,berthier3}
We use an estimate of $T_\mathrm{sg}=1.18(1)$ obtained in our previous
paper.\cite{nakamuraSGzT}
Here, $\tau$ is an only scaling parameter.
From the logarithmic slope we obtain the dynamic critical exponent at the
transition temperature, $z_\mathrm{sg}$. 
When $T<T_\mathrm{cr}$, a typical time, $\tau$, depends on the temperature 
and the dynamic exponent, $z$, is a constant value.
We suppose an activation law for $\tau$ as $\tau=\tau_0\exp[\Delta/T]$,
where $\Delta$ is an average energy barrier of the valley.
In this scaling, $\Delta$ is a scaling parameter.

Figure \ref{fig:isingxi}(b) shows results of the scaling analysis.
The high-temperature scaling is achieved by choosing $\tau=4$.
The results are plotted by symbols.
Each temperature data fall on a single line.
From the slope we obtain $z_\mathrm{sg}=5.1(1)$.\cite{nakamuraSGzT}
The low-temperature scaling is achieved by choosing $\Delta=4J$.
Algebraic increases after the plateau ride on a single line.
The dynamic exponent of this process is estimated to be $z_0=13(1)$.

A value of $\Delta=4J$ is a special number.
It is an energy difference of a local single-spin flip when two spins are 
energetically favorable and four spins are energetically unfavorable among
the six nearest neighbors.
This energy valley is due to a local single-spin flip.
It is not a global excitation valley.
The temperature only controls this local activation.
If we renormalize a Monte Carlo step by the activation time,
we can neglect a dynamic process of trapping by the local energy minimum.
Then, the temperature-independent
ground-state relaxation of the spin-glass correlation length is observed.
It is an averaged global increase of the frozen domain.
The estimated $z_0$ is the ground-state dynamic exponent.

\begin{figure}[t]
\begin{center}
\includegraphics[width=7cm]{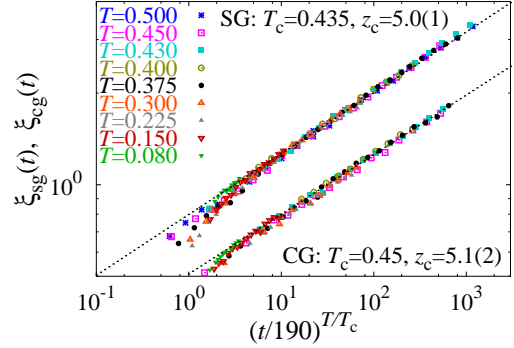}
\end{center}
\caption{(Color online)
Dynamic scaling plots of the SG and the CG correlation lengths
in the {\it XY} model.
CG data are scaled only below $T_\mathrm{sg}$.
}
\label{fig:xyxi}
\end{figure}
Figures \ref{fig:xyxi} shows the dynamic scaling result of the
SG and the CG correlation lengths in the {\it XY} model.
A scaling of $z(T)\propto 1/T$ is supposed.
We use estimates of transition temperatures obtained in our previous
paper.\cite{nakamuraSGzT}
The scaling plot of CG is possible only below $T_\mathrm{sg}$.
The correlation times of SG and CG are same.
The dynamic critical exponent at each transition temperature 
is also consistent with each other.
Therefore, the dynamic behavior in the low-temperature phase is explained by a
critical dynamics of the spin-glass transition.
It is considered to continue to the ground state.
There is no crossover temperature.
Of course, there is a 
possibility of a crossover below the lowest temperature
we have investigated.

\begin{figure}[t]
\begin{center}
\includegraphics[width=7cm]{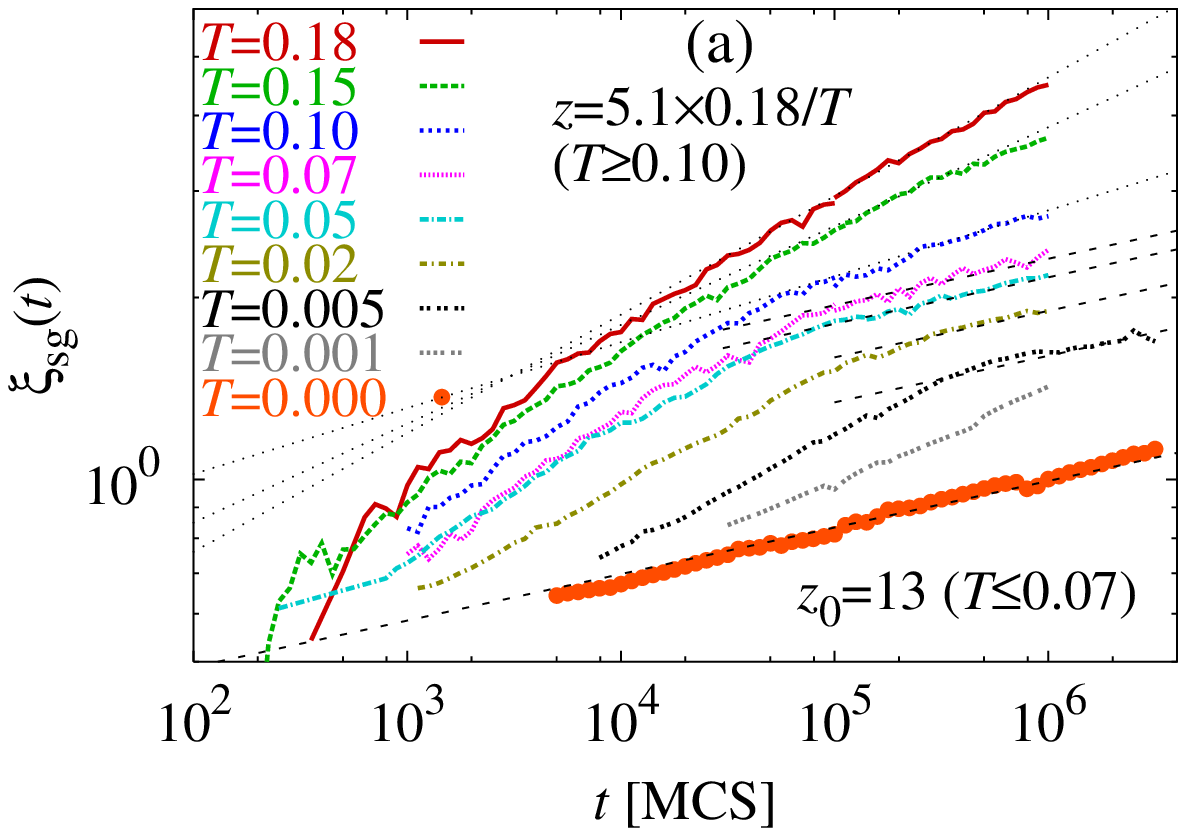}
\includegraphics[width=7cm]{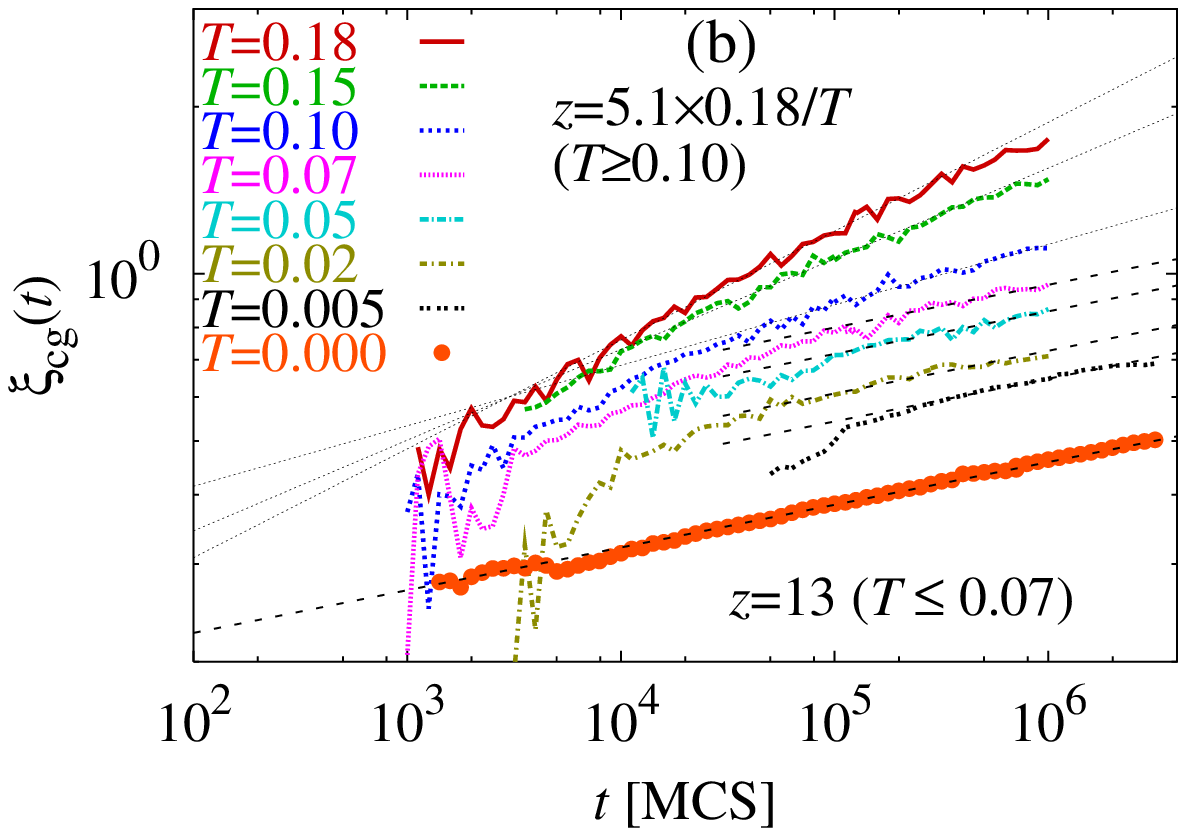}
\includegraphics[width=7cm]{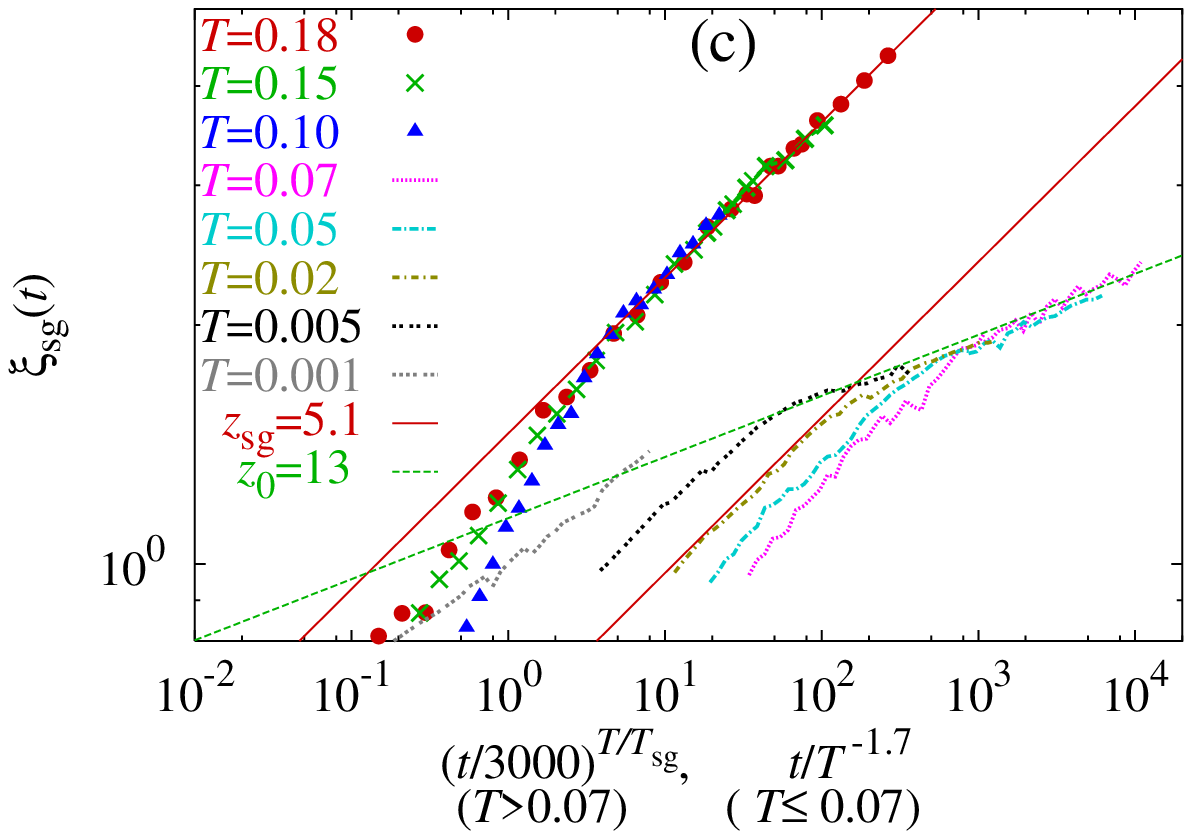}
\end{center}
\caption{(Color online)
Nonequilibrium relaxation functions of the SG(a) and the CG(b) correlation
lengths in the Heisenberg model.
(c) The dynamic scaling plots.
When $T>T_\mathrm{cr}=0.07$, data are plotted with symbols against
$(t/3000)^{T/T_\mathrm{sg}}$.
Below $T_\mathrm{cr}$, they are plotted with lines against 
$t/T^{1.7}$.
Algebraic divergence with $z_\mathrm{sg}$ and that with 
$z_0$ are also plotted for guide for eyes.
}
\label{fig:Hsgxi}
\end{figure}

Figure \ref{fig:Hsgxi} shows the results in the Heisenberg model.
Relaxation functions of the correlation lengths show the same behavior as
in the Ising model.
The logarithmic slope depends on the temperature when $T>T_\mathrm{cr}\simeq
0.07$. 
We have also performed a simulation at $T=0$, which is just a spin-quenching
dynamics.
The correlation length shows algebraic divergence with 
$z_0=13$, which is the ground-state dynamic exponent.
This value coincides with the Ising value.
When $T<T_\mathrm{cr}$, the correlation length exhibits algebraic
divergence with the ground-state dynamic exponent in the long-time limit.
There is no qualitative difference between the spin and the chirality.

A clear difference between the Ising model and the Heisenberg model
is a lack of plateau in the low-temperature relaxation.
The correlation length increases algebraically before it shows the 
ground-state relaxation.
It is considered as a consequence of continuous distribution of the
energy barrier in the local spin flips.
Typical length scale where the ground-state relaxation appears 
depends on the temperature.
There is finite thermal fluctuation even in the spin-glass phase.
It affects the short-range spin correlations.
The ground-state behavior is only observed when the spin-glass correlation 
length exceeds this thermally-fluctuating domain size, which becomes
large as the temperature increases.
In the Ising model, the domain size is considered to be constant due to 
a discrete energy barrier.

The scenario mentioned above is well-depicted 
in the dynamic scaling result in Fig.~\ref{fig:Hsgxi}(c).
We have plotted the spin-glass correlation length against the
renormalized Monte Carlo steps.
When $T>T_\mathrm{cr}$, the scaling analysis is same as the Ising model. 
Typical time, $\tau=3000$, is constant and the dynamic exponent depends
on the temperature as $z(T)=5.1\times T_\mathrm{sg}/T$.
A value of $T_\mathrm{sg}=0.18$ is taken from our previous 
paper.\cite{nakamuraSGzT}
When $T<T_\mathrm{cr}$, the relaxation function is scaled with
the typical time which algebraically diverges as  $T^{-1.7}$.
It is also considered to be a consequence of continuous distribution of
the energy barrier $\Delta$.
We observe the ground-state dynamics after renormalizing the Monte Carlo
steps by this typical time.
The convergence to the ground-state relaxation becomes later as the 
temperature increases.
It is also noted that the relaxation before the convergence is
the critical relaxation with the dynamic exponent $z_\mathrm{sg}$.
Therefore, what is observed in this dynamic scaling is a crossover
from the critical dynamics in the short-time regime to the 
ground-state dynamics in the long-time regime.

\begin{figure}[t]
\begin{center}
\includegraphics[width=6cm]{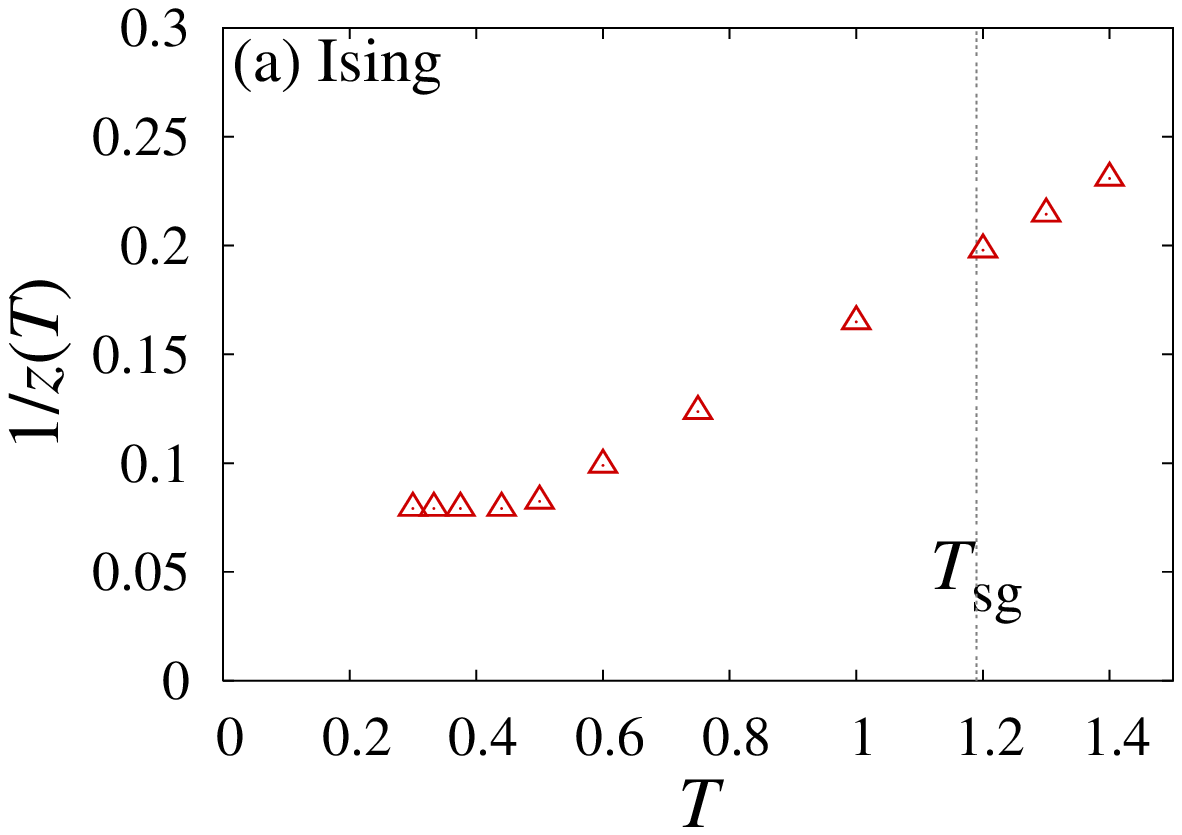}
\includegraphics[width=6cm]{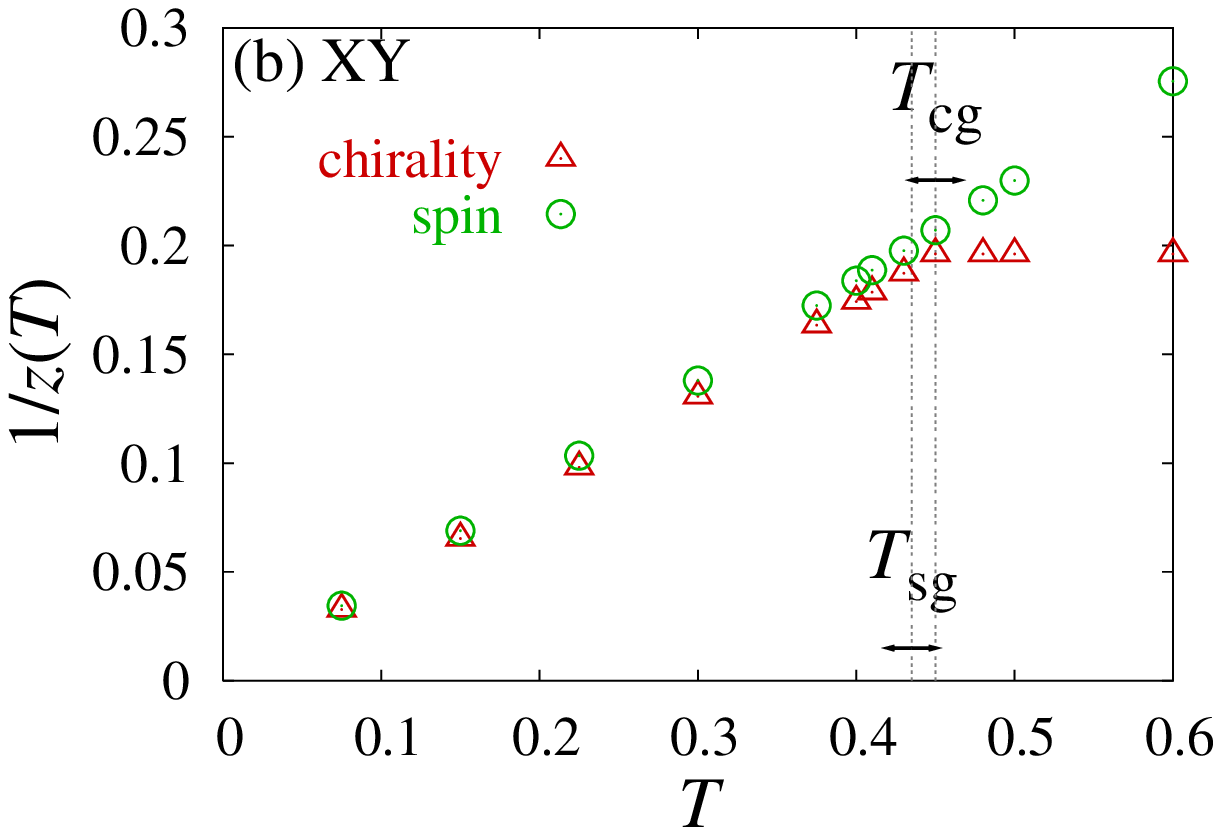}
\includegraphics[width=6cm]{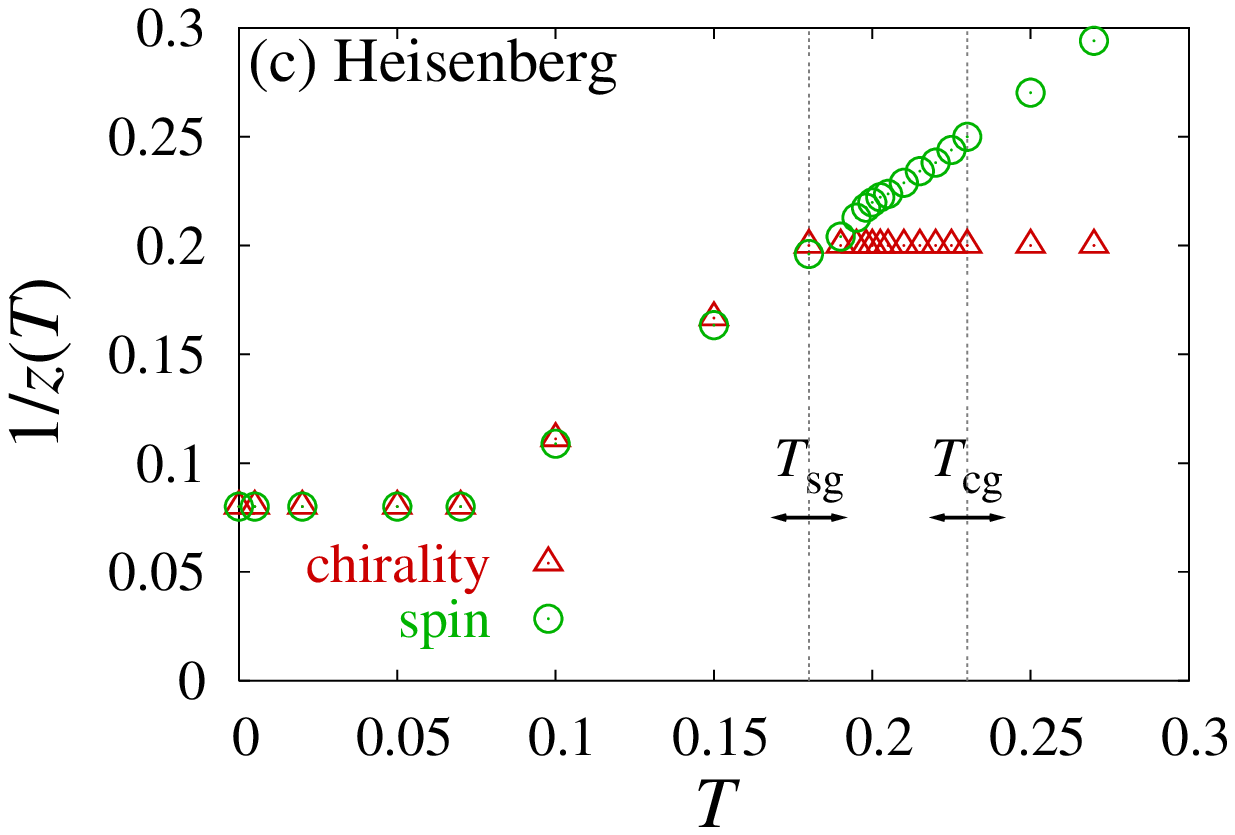}
\end{center}
\caption{(Color online)
Temperature dependences of the nonequilibrium dynamic exponent, $z(T)$, and
the static exponent, $\gamma/\nu$, in the Ising model.
}
\label{fig:isingexponent}
\end{figure}

Figure \ref{fig:isingexponent} summarizes the temperature dependence of the
dynamic exponent.
The Ising model and the Heisenberg model exhibit the same behavior.
The value of the ground-state dynamic exponent agrees with each other.
The similarity suggests that the $Z_2$ symmetry of both models is 
responsible to the dynamic crossover behavior.
We may consider that there is no crossover in the {\it XY} model
due to a lack of this symmetry.

To conclude,
we have investigated the dynamic behavior in the spin-glass phase.
We observed a crossover from the critical dynamics to the 
ground-state dynamics in the Ising model and in the Heisenberg 
model. 
The dynamic exponent $z(T)$ changes the temperature dependence.
The algebraic ground-state dynamics is observed
after renormalizing Monte Carlo steps by a typical time.
The ground-state dynamic exponent is estimated to be $z_0\simeq 13$,
which is common to the two models.
The typical time diverges exponentially in the Ising model,
and it diverges algebraically in the Heisenberg model.
There is no crossover in the {\it XY} model.
The dynamic property of the low-temperature phase is
explained solely by the critical dynamics.

A motivation of this study is to answer a question in regard to 
the nature of the spin-glass phase: RSB or droplet.
The present results alone cannot give an exclusive answer.
However, it is possible to interpret our results by two theories.
In a viewpoint of the droplet theory, 
a crossover from the critical dynamics to the ground-state dynamics is
observed.
Even at very low temperatures, the spin-glass correlation length 
diverges algebraically with time.
It may be the $T=0$ criticality of the droplet theory.\cite{droplet}
However, the ground-state dynamic exponent is estimated to be finite
and is independent of the temperature.
This result is different from the original droplet theory.\cite{droplet}
In the {\it XY} model,
the dynamic exponent seems to diverge at $T=0$.
However, it is difficult to ensure the criticality.
A diverging dynamic exponent suggests that the SG/CG correlation length
does not diverge or at most logarithmically diverges.
In a viewpoint of the RSB theory, 
the present simulation time is too short to observe the RSB behavior.
Only an effect of a local energy valley is observed within our 
algorithm and computer facility.
Up to the present, our results are better explained by the droplet theory.

A dynamic crossover observed in this paper suggests that the spin-glass
ground-state is accessible even at a finite temperature if below the
crossover temperature.
Particularly, the finite energy barrier in the Ising model may be overcome
by a finite quantum fluctuation of the transverse field.
It can be a reason why the quantum annealing method
works well in the spin-glass problems.\cite{qaneal}

The author would like to thank
Professor Nobuyasu Ito and Professor Yasumasa Kanada 
for providing him with a random number generator RNDTIK.
He also thank Dr. Hiroshi Watanabe for a fruitful discussion.
This work is supported by Grant-in-Aid for Scientific Research from
the Ministry of Education, Culture, Sports, Science and Technology, Japan
 (No. 15540358).

\end{document}